\PassOptionsToPackage{unicode}{hyperref}
\PassOptionsToPackage{hyphens}{url}
\PassOptionsToPackage{dvipsnames,svgnames,x11names}{xcolor}
\documentclass[
]{article}
\usepackage{amsmath,amssymb}
\usepackage{lmodern}
\usepackage{iftex}
\ifPDFTeX
  \usepackage[T1]{fontenc}
  \usepackage[utf8]{inputenc}
  \usepackage{textcomp} 
\else 
  \usepackage{unicode-math}
  \defaultfontfeatures{Scale=MatchLowercase}
  \defaultfontfeatures[\rmfamily]{Ligatures=TeX,Scale=1}
\fi
\IfFileExists{upquote.sty}{\usepackage{upquote}}{}
\IfFileExists{microtype.sty}{
  \usepackage[]{microtype}
  \UseMicrotypeSet[protrusion]{basicmath} 
}{}
\makeatletter
\@ifundefined{KOMAClassName}{
  \IfFileExists{parskip.sty}{%
    \usepackage{parskip}
  }{
    \setlength{\parindent}{0pt}
    \setlength{\parskip}{6pt plus 2pt minus 1pt}}
}{
  \KOMAoptions{parskip=half}}
\makeatother
\usepackage{xcolor}
\usepackage{graphicx}
\makeatletter
\def\maxwidth{\ifdim\Gin@nat@width>\linewidth\linewidth\else\Gin@nat@width\fi}
\def\maxheight{\ifdim\Gin@nat@height>\textheight\textheight\else\Gin@nat@height\fi}
\makeatother
\setkeys{Gin}{width=\maxwidth,height=\maxheight,keepaspectratio}
\makeatletter
\def\fps@figure{htbp}
\makeatother
\providecommand{\tightlist}{%
  \setlength{\itemsep}{0pt}\setlength{\parskip}{0pt}}
\NewDocumentCommand\citeproctext{}{}
\NewDocumentCommand\citeproc{mm}{%
  \begingroup\def\citeproctext{#2}\cite{#1}\endgroup}
\makeatletter
 \let\@cite@ofmt\@firstofone
 \def\@biblabel#1{}
 \def\@cite#1#2{{#1\if@tempswa , #2\fi}}
\makeatother
\newlength{\cslhangindent}
\newlength{\csllabelwidth}
\newenvironment{CSLReferences}[2] 
 {\begin{list}{}{%
  \setlength{\itemindent}{0pt}
  \setlength{\leftmargin}{0pt}
  \setlength{\parsep}{0pt}
  \ifodd #1
   \setlength{\leftmargin}{\cslhangindent}
   \setlength{\itemindent}{-1\cslhangindent}
  \fi
  \setlength{\itemsep}{#2\baselineskip}}}
 {\end{list}}
\usepackage{calc}

\ifLuaTeX
\usepackage[bidi=basic]{babel}
\else
\usepackage[bidi=default]{babel}
\fi
\babelprovide[main,import]{american}

\def\languageshorthands#1{}
\ifLuaTeX
  \usepackage{selnolig}  
\fi
\IfFileExists{bookmark.sty}{\usepackage{bookmark}}{\usepackage{hyperref}}
\IfFileExists{xurl.sty}{\usepackage{xurl}}{} 
\hypersetup{
  pdftitle={Foam: A Python package for forward asteroseismic modelling
of gravity modes},
  pdfauthor={Mathias Michielsen},
  pdflang={en-US},
  colorlinks=true,
  linkcolor={Maroon},
  filecolor={Maroon},
  citecolor={Blue},
  urlcolor={Blue},
  pdfcreator={LaTeX via pandoc}}

\title{Foam: A Python package for forward asteroseismic modelling of
gravity modes}

\usepackage[affil-it]{authblk}
\usepackage{orcidlink}
\author[1%
  ]{Mathias Michielsen%
    \,\orcidlink{0000-0001-9097-3655}\,%
    }

\affil[1]{Institute of Astronomy, KU Leuven, Celestijnenlaan 200D,
B-3001 Leuven, Belgium}
\date{1 June 2024}

\begin{document}
\maketitle

\section{Summary}\label{summary}

Asteroseismology, the study of stellar pulsations, offers insights into
the internal structures and evolution of stars. Analysing the variations
in a star's brightness allows the determination of fundamental
properties such as mass, radius, age, and chemical composition.
Asteroseismology heavily relies on computational tools, but a
significant number of them are closed-source, thus inaccessible to the
broader astronomic community. This manuscript presents \texttt{Foam}, a
Python package designed to perform forward asteroseismic modelling of
stars exhibiting gravity modes. It automates and streamlines a
considerable fraction of the modelling process, comparing grids of
theoretical stellar models and their oscillation frequencies to observed
frequency sets in stars.

\texttt{Foam} offers the flexibility to employ diverse modelling
approaches, allowing users to choose different methodologies for
matching theoretically predicted oscillations to observations. It
provides options to utilise various sets of observables for comparison
with their theoretical counterparts, employ different merit functions
for assessing goodness of fit, and to incorporate nested subgrids in a
statistically rigorous manner. For applications of these methodologies
in modelling observed gravity modes, refer to Michielsen et al.
(\citeproc{ref-Michielsen2021}{2021}) and Michielsen et al.
(\citeproc{ref-Michielsen2023}{2023}).

\section{Introduction}\label{introduction}

Stars spend approximately 90\% of their evolution on their so called
\emph{main sequence}, during which they fuse hydrogen into helium in
their cores. In stars with masses above about 1.2 times the mass of the
sun, the stellar core in which these fusion processes take place becomes
convective. Macroscopic element transport in and near the convective
cores of these stars has a large influence on their life, since it
transports additional hydrogen from outside of the nuclear fusion region
into this region. In this way it both prolongs the main-sequence
lifetime of stars and enlarges the mass of the helium core at the end of
the main sequence, which significantly influences all later stages of
their evolution. However, these transport processes provide the largest
uncertainties in stellar structure and evolution models for stars with
convective cores, due to our poor understanding of macroscopic element
transport and limited number of useful observations to test the
theories. (See e.g. \citeproc{ref-Anders2023}{Anders \& Pedersen, 2023}
for a review on this topic.)

Through asteroseismology, we gain the means to unravel the interior
structure of stars (\citeproc{ref-Aerts2010}{Aerts et al., 2010};
\citeproc{ref-Aerts2021}{Aerts, 2021}). Gravity (g-) modes in particular
have a high sensitivity to the properties of the near-core region. These
modes have buoyancy as their restoring force, have dominantly horizontal
displacements, and oscillate with a period of several hours to a few
days. Additionally, they can only propagate in the non-convective
regions in the star, which makes their propagation cavity very sensitive
to the size of the convective core. We can exploit the probing power of
g-modes, observed in e.g.~Slowly Pulsating B-type stars
(\citeproc{ref-Waelkens1991}{Waelkens, 1991}), to investigate the
physics in the interior of these stars, particularly the transition
region between the convective core and radiative envelope.

\section{Statement of need}\label{statement-of-need}

Some tools have been developed and made publicly available to model and
determine stellar parameters of solar-like oscillators, such as
\texttt{AIMS} (\citeproc{ref-Rendle2019}{Rendle et al., 2019}),
\texttt{BASTA} (\citeproc{ref-AguirreBorsenKoch2022}{Aguirre Børsen-Koch
et al., 2022}), and \texttt{pySYD} (\citeproc{ref-Chontos2022}{Chontos
et al., 2022}). However, there are several key differences between the
modelling of the pressure (p-) modes observed in solar-like oscillators,
and the modelling of the g-modes observed in more massive stars. First
and foremost, the well-known asteroseismic scaling relations used for
solar-like oscillators cannot be extrapolated to main-sequence stars
with a convective core. Secondly, the effect of rotation on p-modes is
often included in a perturbative way, whereas the g-mode frequencies are
strongly dependent on rotation and require the inclusion of the Coriolis
acceleration in a non-perturbative way. Additionally the mass regime of
stars with convective cores is subject to strong correlations between
several model parameters, which sometimes follow non-linear
relationships. In this context, the Mahalanobis distance (MD) (see
\citeproc{ref-Aerts2018}{Aerts et al., 2018} for its application to
asteroseismic modelling) provides a more appropriate merit function than
the often used \(\chi^2\), since it tackles both these non-linear
correlations and includes uncertainties for the theoretical predictions.
The use of a different, more appropriate merit function significantly
impacts modelling results. This is demonstrated by Michielsen et al.
(\citeproc{ref-Michielsen2021}{2021}) in their comparison between the
results obtained by employing the MD versus \(\chi^2\), applied in the
modelling of an observed star.

\texttt{Foam} was developed to be complimentary to the available
modelling tools for solar-like oscillators. It provides a framework for
the forward modelling of g-modes in main-sequence stars with convective
cores, and tackles the differences in the modelling approach as compared
to the case of solar-like oscillators. \texttt{Foam} therefore extends
the efforts to provide publicly available, open-source tools for
asteroseismic modelling to the g-mode domain, given that the currently
available tools predominantly concern the solar-like oscillators.

\section{Software package overview}\label{software-package-overview}

\texttt{Foam} is designed as a customisable pipeline. It will match
theoretical models to observations, computing the goodness of fit of
each model based on the selected merit function. Afterwards it will
determine the best model alongside the uncertainty region of this
solution based on statistical criteria. On the observational side, it
will take a list of frequencies as an input, optionally complemented by
additional information such as a set of surface properties (effective
temperature, surface gravity, luminosity, element surface
abundances\ldots). On the theoretical side \texttt{Foam} will use a grid
of theoretical stellar models, calculated by the user to suit their
specific needs. Although the current implementation is made for a grid
of stellar equilibrium models computed by \texttt{MESA}
(\citeproc{ref-Jermyn2023}{Jermyn et al., 2023};
\citeproc{ref-Paxton2011}{Paxton et al., 2011},
\citeproc{ref-Paxton2013}{2013}, \citeproc{ref-Paxton2015}{2015},
\citeproc{ref-Paxton2018}{2018}, \citeproc{ref-Paxton2019}{2019}), whose
pulsation frequencies are computed with \texttt{GYRE}
(\citeproc{ref-Townsend2018}{Townsend et al., 2018};
\citeproc{ref-Townsend2013}{Townsend \& Teitler, 2013}), the majority of
the code is not inherently dependent on \texttt{MESA}. By making certain
adjustments to the modelling pipeline, \texttt{Foam} could potentially
employ grids generated by different stellar evolution codes. Some
suggestions how to approach this are given in the description of
\href{https://michielsenm.github.io/FOAM/Walkthrough}{the theoretical
model grid} in the online documentation. However, the implementation of
such functionality currently remains out of the scope of the project.

The script to run the pipeline can be altered in order to change the
modelling approach you want to take. The various configuration options,
the installation procedure, and a walkthrough of how to create your own
modelling setup, are described in more detail in the
\href{https://michielsenm.github.io/FOAM}{online documentation}.
Although it relies on grids of stellar equilibrium models computed by
\texttt{MESA} as the source of the theoretical model grid,
\texttt{MESA}'s installation itself is not required for \texttt{Foam} to
function. The installation of \texttt{GYRE} is however required,
specifically since \texttt{Foam} relies on the
\texttt{tar\_fit.mX.kX.h5} files included in the \texttt{GYRE}
installation. This allows us to rescale the g-modes for various stellar
rotation rates, following the traditional approximation of rotation
(e.g. \citeproc{ref-Eckart1960}{Eckart, 1960}; see
\citeproc{ref-Townsend2020}{Townsend, 2020} for its implementation in
\texttt{GYRE}) and assuming rigid rotation. This facilitates computing
the oscillation frequencies for the grid of stellar equilibrium models
only once, and subsequently rescaling them to find the optimised
rotation rate (see \citeproc{ref-Michielsen2023}{Michielsen et al.,
2023}). This approach avoids repeating the oscillation computations for
a variety of rotation values, which would introduce extra dimensionality
in the modelling problem in the form of adding the rotation rate as an
additional free parameter.

\begin{figure}
\centering
\includegraphics{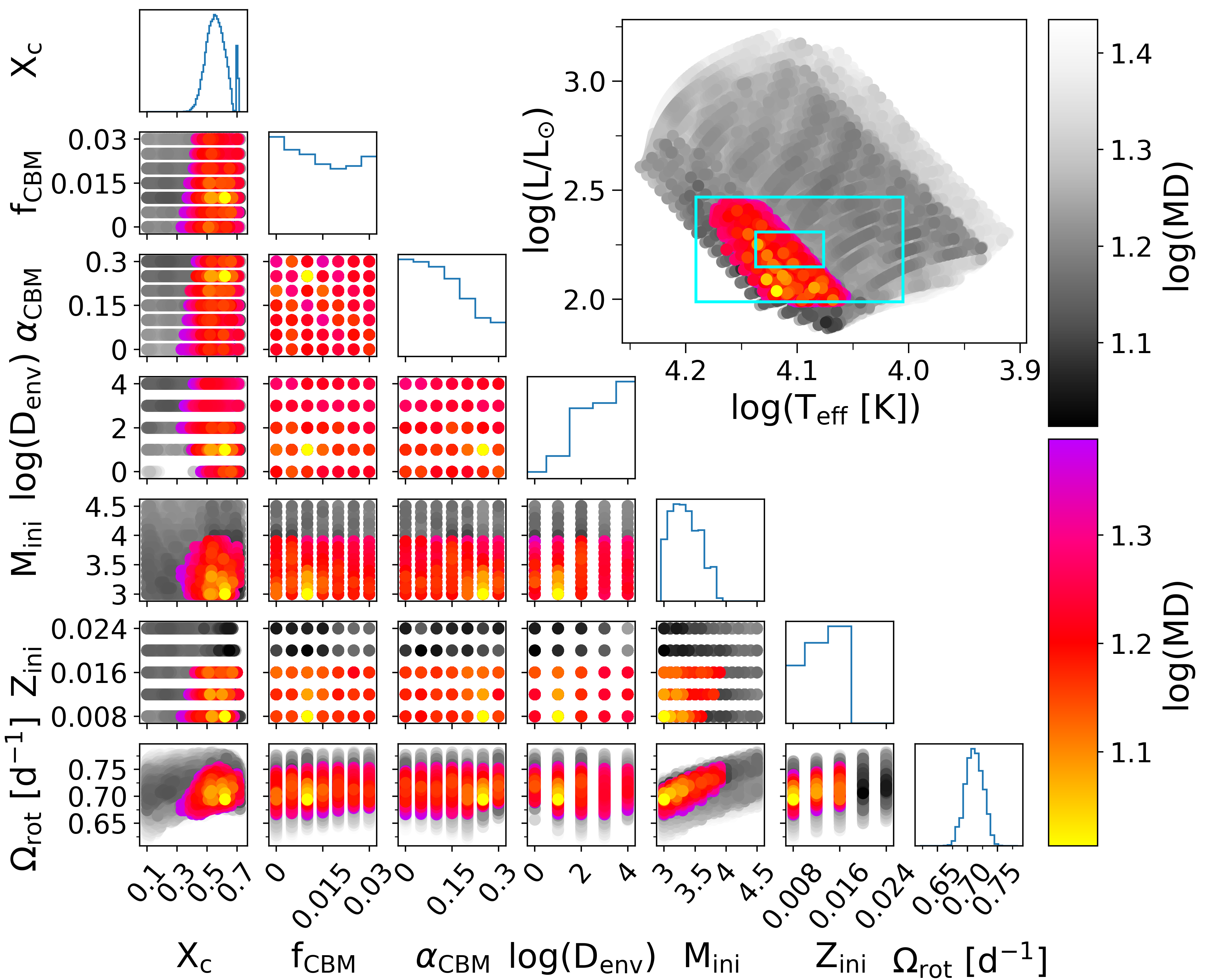}
\caption{Cornerplot with the parameters in the grid and the rotation.
The 50\% best models are shown, colour-coded according to the log of
their merit function value. Models in colour fall within the 2 sigma
error ellipse, while those in greyscale fall outside of it. Figures on
the diagonal show binned parameter distributions of the models in the
error ellipse, and the panel at the top right shows an
Hertzsprung-Russell (HR) diagram with 1 and 3 sigma observational error
boxes. Figure taken from Michielsen et al.
(\citeproc{ref-Michielsen2023}{2023}). \label{fig:cornerplot}}
\end{figure}

\texttt{Foam}'s modelling procedure can be broken down into following
sequential \href{https://michielsenm.github.io/FOAM/Pipeline}{steps of
the pipeline}:

\begin{itemize}
\tightlist
\item
  Extract all required parameters and quantities from the files in the
  theoretical \texttt{MESA} and \texttt{GYRE} grids.
\item
  Construct the theoretical pulsation patterns for each stellar model.
  Thereafter select theoretical pulsation patterns matching the
  observational pattern whilst optimising their rotation rates. Finally
  combine this information with the models' surface properties.
\item
  Calculate the likelihood of all the theoretical patterns according to
  the specified merit functions and observables. This list of
  observables consist of the pulsations, but can optionally be extended
  with spectroscopic or astrometric information.
\item
  Exclude all the models that fall outside an n-sigma error box on the
  spectroscopic and astrometric constraints as acceptable solutions.
\item
  Calculate the Akaike information criterion (AIC,
  \citeproc{ref-Claeskens2008}{Claeskens \& Hjort, 2008}) corrected for
  small sample size. This statistical criterion rewards goodness of fit,
  but penalises model complexity in the form of additional free
  parameters. The AIC thus allows a statistical comparison between
  models of different (nested) grids where the number of free parameters
  is not the same.
\item
  Calculate the 2 sigma uncertainty region of the maximum likelihood
  solution using Bayes' theorem.
\item
  Make corner plots for all combinations of the different modelling
  choices (See \autoref{fig:cornerplot} for an example).
\item
  Construct a table with the best model of the grid for each combination
  of different modelling choices.
\end{itemize}

Next to the tables with the best model parameters and their AIC values,
the cornerplots provide a quick way to assess the output of the pipeline
and visualise the modelling results. \autoref{fig:cornerplot} shows an
example of such a cornerplot for the modelling of KIC 4930889 performed
by Michielsen et al. (\citeproc{ref-Michielsen2023}{2023}). It gives a
clear indication of which models are included (coloured) or excluded
(greyscale) from the uncertainty region, and indicates what the best
models of the grid are (yellow, see the colour bar).

\section{Acknowledgements}\label{acknowledgements}

The research leading to the development of this package has received
funding from the Research Foundation Flanders (FWO) by means of a PhD
scholarship to MM under project No.~11F7120N. MM is grateful to Timothy
Van Reeth for his help concerning the scaling of g-modes with rotation,
to Alex Kemp for his suggestions regarding the online documentation, and
to the reviewers Ashley Chontos and Ankit Barik for their constructive
remarks.

\section*{References}\label{references}
\addcontentsline{toc}{section}{References}

\phantomsection\label{refs}

\end{document}